\documentclass[
 reprint,
 amsmath,
 pre,
 amssymb,
 nofootinbib,
 aps,
]{revtex4-1}

\usepackage[pdftex]{graphicx}
\usepackage{wasysym}
\usepackage{amsfonts}
\usepackage{mathtools}
\usepackage{stackengine}
\usepackage{dsfont}
\usepackage{ifsym}
\usepackage{slashed}

\makeatletter
\newlength{\apb@width}
\newcommand{\autoparbox}[2][c]{\settowidth{\apb@width}{#2}\parbox[#1]{\apb@width}{#2}}

\makeatother

\newcommand{\namedref}[2]{\hyperref[#2]{#1~\ref*{#2}}}

\newcommand{\Csphere}{{}^\bullet\kern-1.2pt C}
\newcommand{\Ctorus}{{}^\circ\kern-1.2pt C}

\newcommand{\nn}{\nonumber}

\newcommand{\COMMENT}[1]{}

\newcommand{\neqa}{\nonumber\end{eqnarray}}
\newcommand{\la}[1]{\label{#1}}

\newcommand{\<}{{\langle}}
\renewcommand{\>}{{\rangle}}

\newcommand{\re}{\relax{\rm I\kern-.18em R}}

\def\su2{{SU(2)}}

\def\[{\left[}
\def\]{\right]}

\def\({\left(}
\def\){\right)}
\def\[{\left[}
\def\]{\right]}

\def\<{\langle}
\def\>{\rangle}

\def\i2{\frac{i}{2}}

\def\2F1{\,_2{\rm F}_1}

\usepackage[usenames,dvipsnames]{xcolor}
\usepackage{color}
\usepackage{graphicx}
\usepackage{dcolumn}
\usepackage{bm}
\usepackage{hyperref}

\usepackage{enumitem} 

\usepackage{cancel}
\usepackage{ctable}
\usepackage{booktabs}
\newcolumntype{L}[1]{>{\raggedright\let\newline\\\arraybackslash\hspace{0pt}}m{#1}}
\newcolumntype{C}[1]{>{\centering\let\newline\\\arraybackslash\hspace{0pt}}m{#1}}
\newcolumntype{R}[1]{>{\raggedleft\let\newline\\\arraybackslash\hspace{0pt}}m{#1}}

\newcommand{\beq}{\begin{equation}}
\newcommand{\eeq}{\end{equation}}
\newcommand{\beqq}{\begin{equation*}}
\newcommand{\eeqq}{\end{equation*}}
\newcommand\beqa{\begin{eqnarray}}
\newcommand\eeqa{\end{eqnarray}}
\newcommand\beqaa{\begin{eqnarray*}}
\newcommand\eeqaa{\end{eqnarray*}}
\newcommand\bea{\begin{array}}
\newcommand\eea{\end{array}}

\begin{document}


\title{SUSY S-matrix Bootstrap and Friends}

\author{Carlos Bercini} 
\affiliation{}
\affiliation{Perimeter Institute for Theoretical Physics, 31 Caroline St N Waterloo, Ontario N2L 2Y5, Canada}
\affiliation{Instituto de F\'isica Te\'orica, UNESP, ICTP South American Institute for Fundamental Research, Rua Dr Bento Teobaldo Ferraz 271, 01140-070, S\~ao Paulo, Brazil}
\author{Matheus Fabri}
\affiliation{}
\affiliation{Perimeter Institute for Theoretical Physics, 31 Caroline St N Waterloo, Ontario N2L 2Y5, Canada}
\affiliation{Instituto de F\'isica Te\'orica, UNESP, ICTP South American Institute for Fundamental Research, Rua Dr Bento Teobaldo Ferraz 271, 01140-070, S\~ao Paulo, Brazil}
\author{Alexandre Homrich}
\affiliation{}
\affiliation{Perimeter Institute for Theoretical Physics, 31 Caroline St N Waterloo, Ontario N2L 2Y5, Canada}
\affiliation{Instituto de F\'isica Te\'orica, UNESP, ICTP South American Institute for Fundamental Research, Rua Dr Bento Teobaldo Ferraz 271, 01140-070, S\~ao Paulo, Brazil}
\author{Pedro Vieira}
\affiliation{}
\affiliation{Perimeter Institute for Theoretical Physics, 31 Caroline St N Waterloo, Ontario N2L 2Y5, Canada}
\affiliation{Instituto de F\'isica Te\'orica, UNESP, ICTP South American Institute for Fundamental Research, Rua Dr Bento Teobaldo Ferraz 271, 01140-070, S\~ao Paulo, Brazil}


\begin{abstract}
We consider the 2D S-matrix bootstrap in the presence of supersymmetry, $\mathbb{Z}_2$ and $\mathbb{Z}_4$ symmetry. At the boundary of the allowed S-matrix space we encounter well known integrable models such as the  supersymmetric sine-Gordon and restricted sine-Gordon models, novel elliptic deformations thereof, as well as a two parameter family of $\mathbb{Z}_4$ elliptic S-matrices previously proposed by Zamolodchikov. We highlight an intricate web of relations between these various S-matrices.  

\end{abstract}

\pacs{Valid PACS appear here}
\maketitle


\section{Some beautiful sections} \label{sec: introduction}

The S-matrix bootstrap aims at determining the space of possible S-matrices in unitary relativistic quantum field theories. The S-matrix space can be very rich, with pointy structures such as edges and cusps. It is hard to visualize it since we are dealing with an infinite dimensional space so in practice we pick sections. If the theory has bound states, for instance, a natural set of variables to follow are the residues of S-matrix elements at their corresponding poles which physically correspond to the on-shell three particle couplings. While if the theory has no bound states we can measure the two-to-two S-matrix elements at some off-shell points, thus defining effective off-shell four point couplings. By picking appropriate linear functionals and S-matrix ansatze, we thus explore the possible S-matrix space sections compatible with crossing and unitarity following \cite{Paper2,Paper4}. 
In this letter we will consider a few simple sections which are two or three dimensional and thus can be nicely plotted. The physical setups we will consider are:
\begin{itemize}
	\item[\bf(A)] Scattering of a massive real supermultiplet with and without bound states.
		\item[\bf(B)] A generic degenerate boson-fermion scattering where the previous case should sit in a special limit.
		\item[\bf(C)] $\mathbb{Z}_4$ symmetric models.
\end{itemize}
We will always be in two spacetime dimensions. 

These examples are richer than the setup of \cite{Paper1,Paper2,Paper3} where the scattering of the lightest real bosonic particles in a gapped theory was considered but still simpler than the scattering of particles in the fundamental representation of a $O(N)$ flavour symmetry  \cite{lucia,Martin}, with $U(N)$ symmetry \cite{UN-Paulos} or when we scatter the two lightest particles in $\mathbb{Z}_2$ symmetric 2D theories \cite{Paper4}. The great merit of the simpler examples considered herein is that they are simple enough to be able to be analytically described while rich enough to capture many of the intricate features of these other more elaborate examples. 
To generate all the plots here we followed the usual numerical algorithms in S-matrix bootstrap explorations, see appendix \ref{appendixNumerics} for a telegraphic summary. 



\subsection{The simplicity of supersymmetry}\
\label{susysetup}

As the first example we consider a system with $\mathcal{N}=1$ supersymmetry in which the lightest supermultiplet consists of a single real boson $\phi$ and a Majorana fermion $\psi$ both of mass $m$. There are five possible two-to-two scattering amplitudes but SUSY relates most of them so that in the end only two channels are independent: the scattering of bosons $S_{\phi\phi}^{\phi\phi}(s)$ and the forward scattering of a boson against a fermion $S_{\phi\psi}^{\phi\psi}(s)$. These two amplitudes are crossing symmetric. They may have poles corresponding to bosonic or fermionic bound-states which would also be in an $\mathcal{N}=1$ multiplet, hence with couplings all related by SUSY, see appendix \ref{SUSYAlgebra} for details. Evaluated at the crossing
symmetric point $s_*=2m^2$ these two amplitudes define a nice two dimensional section of effective four point off-shell couplings which we can use to probe the supersymmetric S-matrix space.

\begin{figure}[t!]
	\centering
	\includegraphics[width=\linewidth]{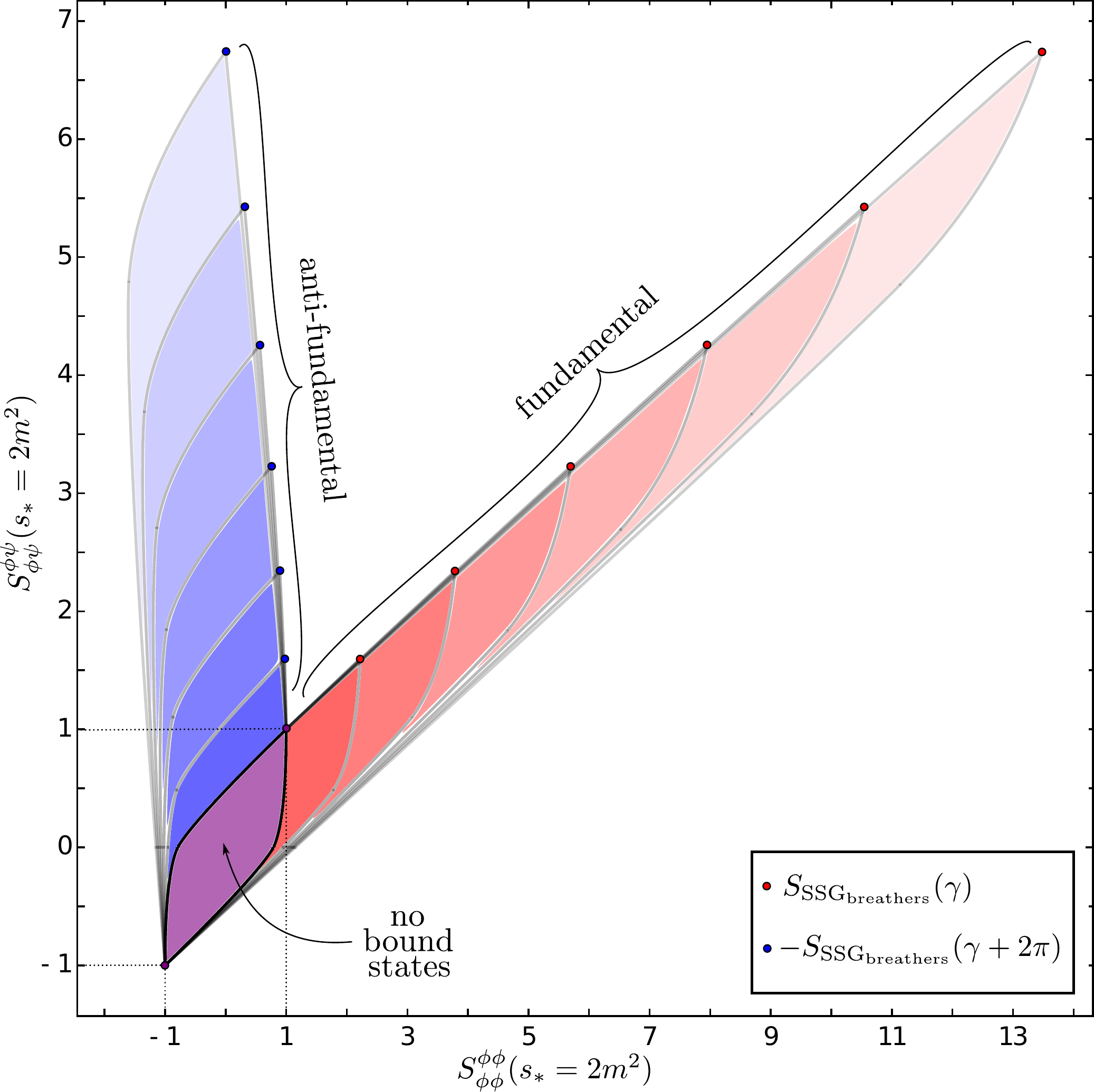}
	\caption{Allowed  $\mathcal{N}=1$ S-matrix space with a single bound state of mass $m_{\text{bs}}/m=\{1.73,1.76,1.80,1.85,1.90,1.96 \}$ transforming in the fundamental/anti-fundamental representation represented in red/blue. As we increase the mass of the bound state the allowed space shrinks. 
}
	\label{fig:masterpiece}
\end{figure}

The space allowed for the two independent  quartic off-shell couplings is depicted in figure \ref{fig:masterpiece}. In purple, the smallest region, corresponds to the allowed coupling space for theories with no bound states. Then the S-matrix elements have no poles inside the physical strip. This purple football-like shape has two cusps corresponding to the free theories where $S= \pm \mathds{I}$. At its boundary we find a remarkable well-known S-matrix: it's nothing but the lightest breather-breather S-matrix of the supersymmetric sine-Gordon theory (SSG) stripped out of the overall CDD-pole. This is also known as the breather S-matrix of the restricted sine-Gordon model (RSG) although this is quite a misnomer since the RSG model has no bound states. For a brief review of the so called RSG model see appendix \ref{rsgappendix}.  The purple shape's boundary can actually be read off from the RSG S-matrices and possess a nice closed form
\beqa
&&\ \ \!\! \left({ S_{\phi\phi}^{\phi\phi}(s_*) },{ S_{\phi\psi}^{\phi\psi}(s_*)}\right)_\texttt{boundary} = \pm(1 \pm 2a,1 )\times \nn \\
&&\times \frac{\exp\! \Big(\frac{i}{\pi} \text{Li}_2\big(\frac{i  (1-2 a \sqrt{1-a^2} )}{2 a^2-1}\big)-\frac{i}{\pi} \text{Li}_2\big(\frac{i  (2 a \sqrt{1-a^2}-1 )}{2
   a^2-1}\big)-\frac{2 \mathcal{C}}{\pi }\Big)}{\sqrt{1-a^2}+a}, \nn
\eeqa
where $\mathcal{C} \simeq 0.915966 $ is the Catalan's constant and~$a>0$. It is quite amusing to see such rich analytic structure arise from such a simple convex optimization problem. From an algebraic perspective, it is quite remarkable that all along the purple region we obtain S-matrices which obey the Yang-Baxter factorization condition although this condition was not imposed in any way. 

%

In addition to the scattered boson and fermion we also consider a setup where there is a single bound state supermultiplet $(b,f)$ of mass $m_{\textrm{bs}}$ with $b$ and $f$ being the bosonic and fermionic bound states, respectively. This is implemented by  allowing for simple poles in the physical sheet in the previous S-matrix elements at $s$ and $t$ equal to $m_{\text{bs}}^2$. As explained in appendix \ref{SUSYAlgebra} the bound state supermultiplet can transforms in the {fundamental} or the {anti-fundamental} representaion. These differ for slightly different relations between the couplings arising in the S-matrix elements, see equations (\ref{eq:SUSYResidues}). The allowed S-matrix space for both cases obtained from the numerical optimization is depicted in figure \ref{fig:masterpiece}. The various red or blue regions correspond to the allowed S-matrix space for various bound state masses $m_\text{bs}$ in either of the two possible representations. As the mass of the bound state increases these regions shrink. When $m_\text{bs}=2m$, the bound state dissolves into the two-particle threshold and we recover the bound state free space depicted in purple at figure \ref{fig:masterpiece}. 

The vertex at the top right corner of the red regions -- corresponding to the S-matrix space with a single fundamental multiplet bound state -- corresponds to the lightest breather S-matrix of the supersymmetric sine-Gordon model (SSG) \cite{ahn}. We could also find that the S-matrix living at the top cusp of the blue regions -- corresponding to the S-matrix space with a single anti-fundamental multiplet bound state -- is an analytic continuation of the SSG S-matrix multiplied by an overall minus sign, see appendix \ref{sec:ssg} for details. We do not know of a Lagrangian theory which realizes this factorized S-matrix. Finally, we have the boundaries connecting to these red and blue vertices. We were able to find the exact S-matrices living at these boundaries, see appendix \ref{nissg}. They saturate unitarity as usual but don't satisfy the Yang-Baxter factorization equations. These S-matrices are most likely not physical S-matrices but perhaps they are close enough to physical S-matrices with very little particle production. Finally, note that all this seems to be consistent with the classical intuition from \cite{Bercini} where it was found that the only supersymmetric model with a single real scalar boson and a Majorana fermion, with a Lagrangian description and without tree level particle production is the SSG model. Would be interesting to see if the blue cusps admits a Lagrangian description in terms of a fermion plus a pseudo-scalar. 

\subsection{How special is SUSY?}
\label{z2sec}
Supersymmetric theories are special instances of theories with bosons and fermions with further non-bosonic symmetries relating them. It is thus natural to look for generic theories with bosons and fermions \textit{without} supersymmetry and see whether supersymmetry, with its extra structure, emerges naturally at special points in the allowed theory space. This is what we turn to next.

\begin{figure}[t!]
	\centering 
	\includegraphics[width=1.05\linewidth]{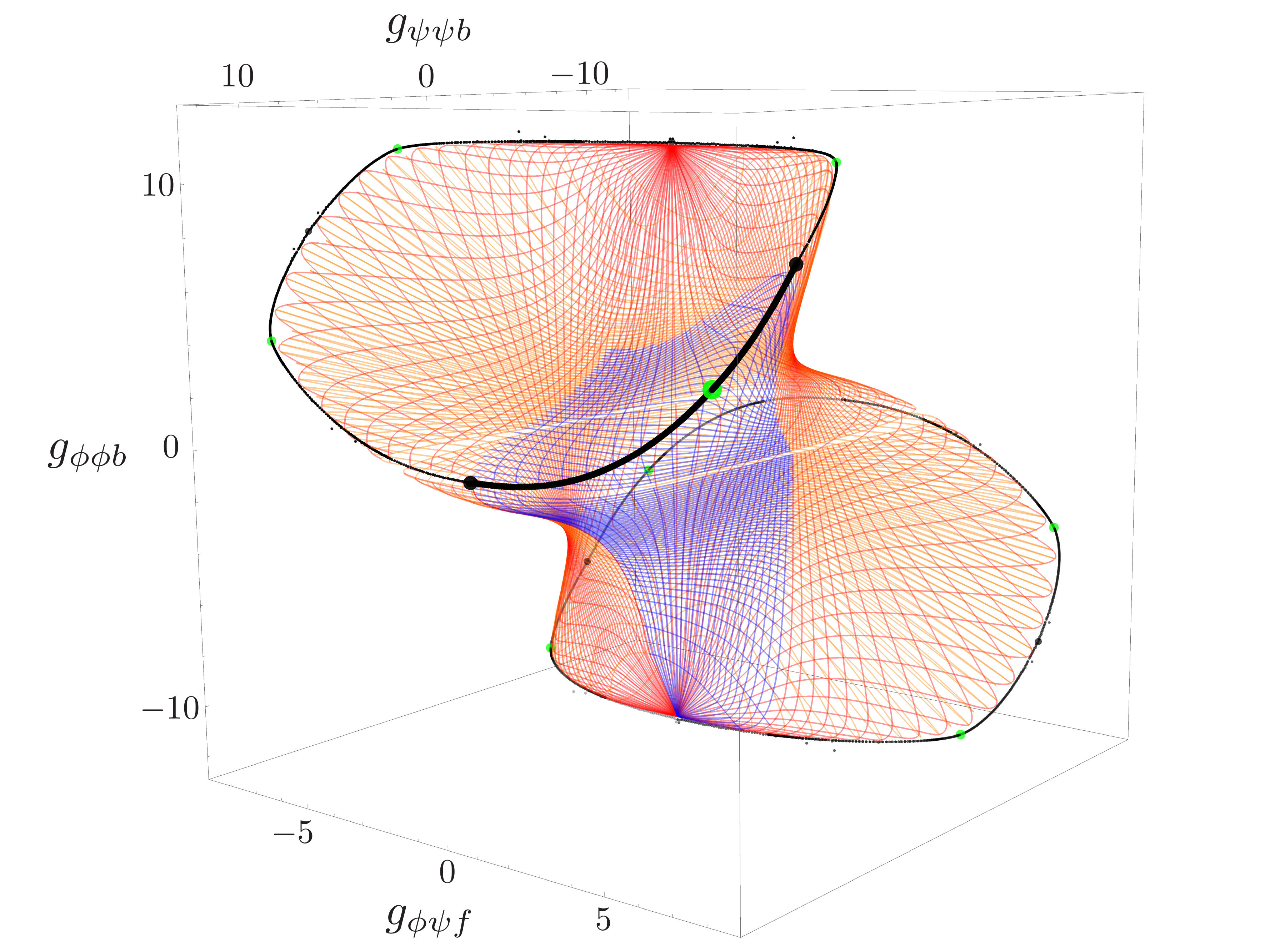}
	\vspace{0.75cm} 
	\caption{Coupling space for $\mathbb{Z}_2$ symmetric theories with a single bound state obtained from the S-matrix bootstrap. 
%
In this figure $m_\text{bs} =\sqrt{3} m$ and all couplings are measured in units of~$m$. The green point is the supersymmetric sine-Gordon theory, while the bold black line corresponds to an integrable elliptic deformation of SSG. The blue region is the fundamental domain: the rest of the 3D space can be obtained from it through trivial reflections corresponding to symmetries of the bootstrap problem.}
	\label{fig: z2plot}
\end{figure}

We consider
a general $\mathbb{Z}_2$ symmetric system with an even (the boson $\phi$) and an odd particle (the fermion $\psi$) with the same mass $m$, but a priori no symmetry relating them. To make contact with the previous bounds we also assume the existence of a boson ($b$), fermion ($f$) pair of bound states both with the same mass $m_\text{bs}$ but, again, with no symmetry relating them. We then have a nice three dimensional section of the allowed S-matrix space parametrized by the three independent couplings $g_{\phi\phi b}$, $g_{\psi\psi b}$ and $g_{\phi\psi f}$. This space can be plotted following \cite{Paper4}; the result is the nice hourglass looking coupling space shown in figure \ref{fig: z2plot}.
The supersymmetric sine-Gordon model beautifully appears as a special point (the green dot) on the boundary of the allowed space. At this point, all three couplings are related by supersymmetry. We also encounter an elliptic deformation of the SSG model (black curve) previously obtained in \cite{Paper4}.\footnote{Strictly speaking the elliptic deformation found in \cite{Paper4} is an analytic continuation of the one found here. Here we are taking $m_\text{bs} > \sqrt{2}m$ to pass by the SSG in its physical domain where the second breathers are constrained to be in such mass range. There we took $m_\text{bs}=1$ so we were instead studying the elliptic deformation of an analytic continuation of the SSG beyond its physical regime. We expect the elliptic deformation encountered here to correspond to a proper physical theory; we suspect that this is not the case for the analytically continued version in \cite{Paper4}.}
This elliptic deformation contains a parameter $\kappa$ and varying it in the allowed range yields the bold curve in figure \ref{fig: z2plot}, in special when $\kappa=0$ we recover SSG. This elliptic deformation preserves integrability, but breaks supersymmetry and its explicit form is given in appendix \ref{sec:ellipticdef}.

This elliptic S-matrix is \textit{not} the famous Zamolodchikov's $\mathbb{Z}_4$ S-matrix found in \cite{z4paper}; the $\mathbb{Z}_2$ S-matrix we found has a different matrix structure and contains a bound-state. Nonetheless, it does share many of its properties. Given that we encounter such rich elliptic solutions at the boundary of the allowed S-matrix space it is most natural to look for Zamolodchikov's $\mathbb{Z}_4$ S-matrix and see if that one can also be found in an appropriate bootstrap problem. This is what we discuss in the next section. 


\subsection{The faces of $\mathbb{Z}_4$ symmetry}

Inspired by the newly obtained elliptic S-matrix discussed in section \ref{z2sec}, we consider a $\mathbb{Z}_4$ symmetric setup with a particle-antiparticle pair of mass $m$ whose charges under $\mathbb{Z}_4$ are one and three. We assume that there are no further particles in the spectrum 

After imposing selection rules from charge conservation and constraints from crossing, C, P and T, see details in appendix \ref{selection}, we are left with 3 independent amplitudes: $S_{11}^{11}$, $S_{11}^{33}$ and $S_{13}^{31}$. 
In similar spirit to the scenario without bound states considered in the SUSY setup, section \ref{susysetup}, we bootstrap the allowed space for the off-shell four point couplings defined by the values of these three independent amplitudes evaluated at the crossing symmetric point $s_* = 2m^2$. 

\begin{figure}[t!]
	\centering 
	\includegraphics[width=0.95\linewidth]{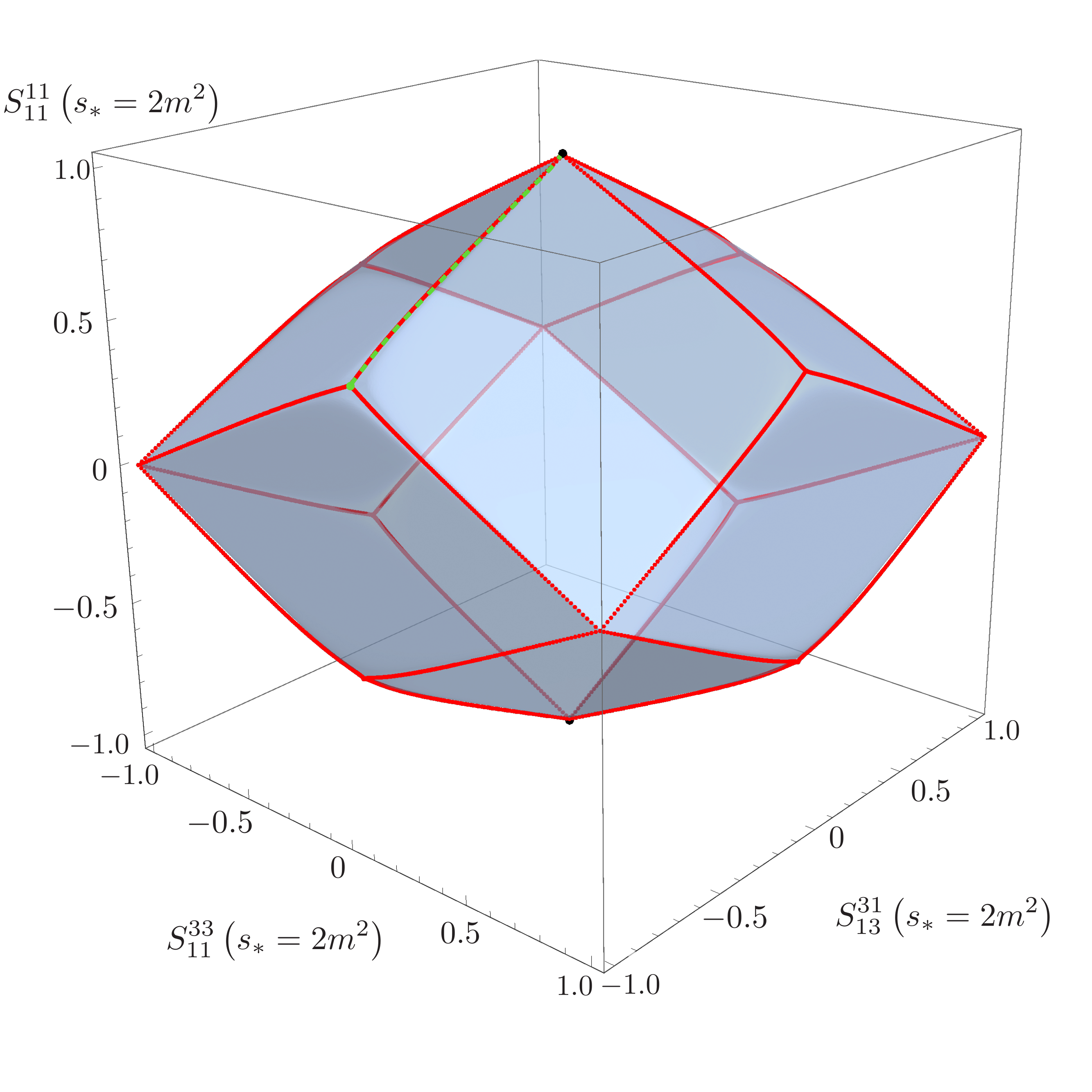}
	\vspace{0.05cm}
	\caption{S-matrix space for the $\mathbb{Z}_4$ symmetric S-matrix bootstrap at $s_*=2m^2$. The faces are equivalent to the Zamolodchikov's $\mathbb{Z}_4$ S-matrices (appendix \ref{sec:exactz4matrix}) and the edges to the sine-Gordon kinks S-matrices (appendix \ref{sgappendix}).}
	\label{z4plot}
\end{figure}

The result is the smoothed rhombic dodecahedron displayed in figure \ref{z4plot}. Yang-Baxter factorization once again makes an unexpected appearance: the \textit{full} two dimensional surface\footnote{More precisely, part of the surface corresponds to Zamolodchikov's $\mathbb{Z}_4$ S-matrix after charge conjugation of one of the particles, see appendix \ref{selection}.} corresponds to Zamolodchikov's famous $\mathbb{Z}_4$ symmetric integrable S-matrix \cite{z4paper}. Edges connecting threefold vertices and fourfold vertices correspond, up to change of basis, to limits where the $\mathbb{Z}_4$ S-matrix degenerates into the sine-Gordon kinks S-matrix with $\gamma \geq \pi$, see section \ref{relations}. In particular, fourfold vertices are equivalent to limits in which the $\mathbb{Z}_4$ or sine-Gordon S-matrix becomes free
. The threefold vertices of the dodecahedron are smoothened resembling the pre-vertices of \cite{DualPaperProblem}.

As far as Yang-Baxter is concerned we encountered this mysterious bonus factorization at special kinks in the supersymmetric setup (figure \ref{fig:masterpiece}); at special lines in the $\mathbb{Z}_2$ bounds (figure \ref{fig: z2plot}) and now in full surfaces in the $\mathbb{Z}_4$ problem (figure \ref{z4plot}). Would be great to understand mathematically where this additional physical factorization is coming from. 

\section{A web of relations}\label{relations}
\begin{figure}[t!]
	\centering 
	\includegraphics[width=\linewidth]{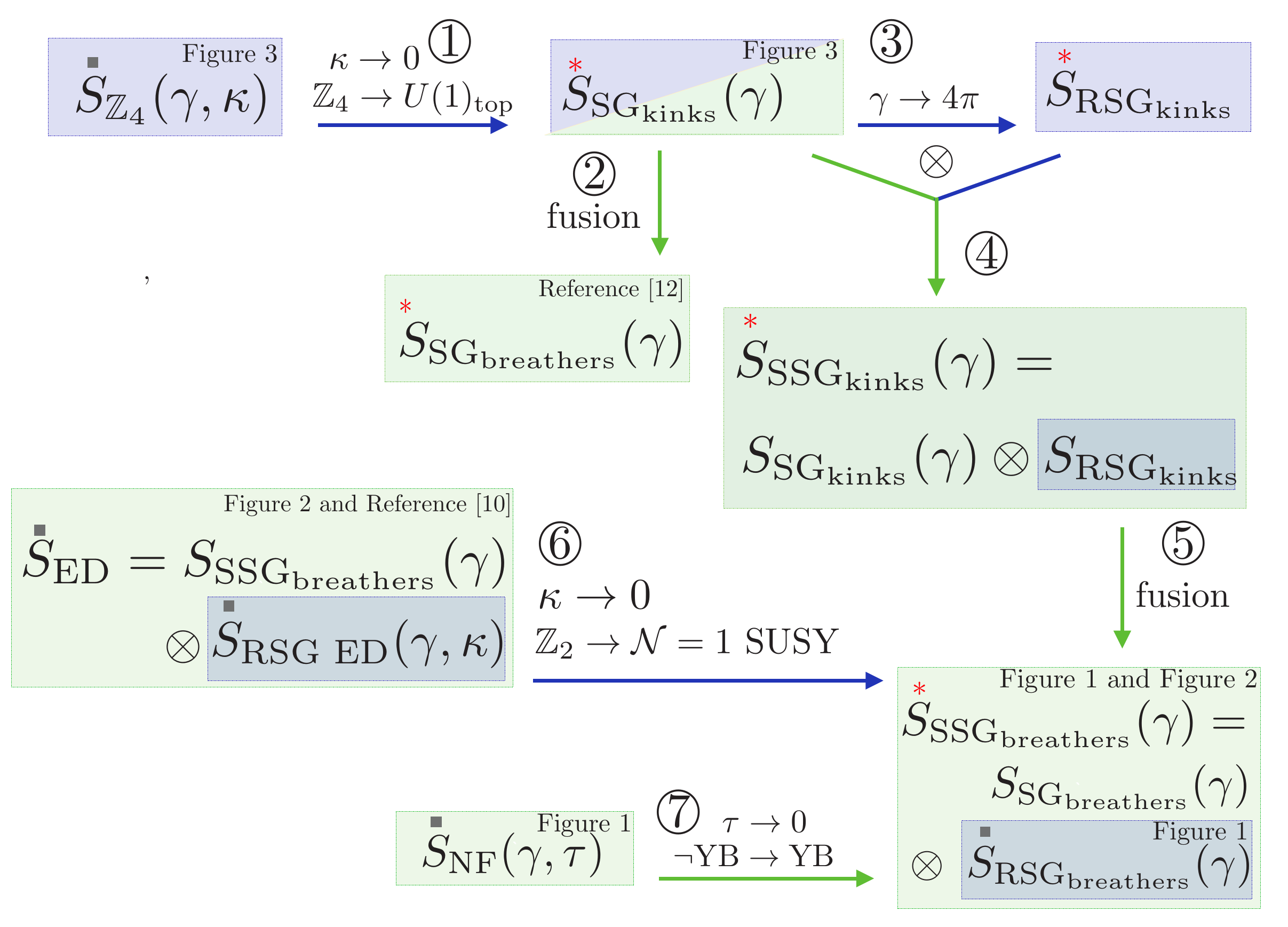}
	\caption{Connections between S-matrices showing up in this work as well as in \cite{Paper2,Paper4}. {\color{ForestGreen}\textit{Green boxes}}: S-matrices with bound-states. {\color{MidnightBlue}\textit{Blue boxes}}: S-matrices without bound states. {\color{Red}$\ast$}: Known corresponding Lagrangian field theory (LFT). {\color{Gray}$\blacksquare$}: Unknown corresponding LFT. }
	\label{web}
\end{figure} 

Both in this and in previews works \cite{Paper2, Paper4} a myriad of integrable two-component 2D S-matrices were found to be located along the boundary of the space of amplitudes allowed by consistency with UV completeness. The various S-matrices obtained in this way are not independent, but connected through an intricate web of relations, summarized in figure \ref{web} and reviewed in this section. The expressions for the exact S-matrices can be found in appendix \ref{exact} where more details are given. 

We begin the web of relations with the Zamolodchikov's $\mathbb{Z}_4$ S-matrix, bootstrapped in figure \ref{z4plot}. The most curious feature of this S-matrix, described in details in appendix \ref{sec:exactz4matrix}, is its periodicity for real values of the rapidity $\theta$, defined by $s = 4m^2 \cosh(\theta/2)^2$, which at high energies amounts to periodicity in $\log s$. As pointed out by Zamolodchikov \cite{z4paper}, this suggests a sort of RG-time periodicity, which may explain the current lacks of a Lagrangian description for this model. The S-matrix is described by two parameters: the elliptic modulus $\kappa$ and the coupling $\gamma$.
When we take $\kappa \rightarrow 0$ (arrow $\raisebox{.5pt}{\textcircled{\raisebox{-.9pt} {1}}}$) the $\mathbb{Z}_4$ charge gets enhanced to a $U(1)$ topological charge, and the S-matrix gets reduced to the sine-Gordon kinks S-matrix. The remaining real parameter is the free parameter $\gamma$ of the sine-Gordon model 
which controls the spectrum of the theory. 


As a limit of the $\mathbb{Z}_4$ S-matrix (which has no bound-states) we land in the regime $\gamma>\pi$ where the only stable particles are the sine-Gordon solitons. Once we analytically continue into $\gamma<\pi$ we reach the regime where there are bound states called breathers. The scattering of these breathers can be obtained by fusing pairs of kinks in a multi-kink scattering process (arrow $\raisebox{.5pt}{\textcircled{\raisebox{-.9pt} {2}}}$), detailed in appendix \ref{sgappendix}. The lightest breather S-matrix, obtained in this way, is the simplest S-matrix one can bootstrap as analyzed in \cite{Paper2,LatticeGuy}.

As said previously, for $\gamma > \pi$ the only stable particles in the sine-Gordon spectrum are the solitons. However when $\gamma=\pi p$, with $p\geq 3$ and $p\in\mathbb{Z}$,
 some multi-soliton states decouple and the spectrum can be restricted (arrow  $\raisebox{.5pt}{\textcircled{\raisebox{-.9pt} {3}}}$). This process defines the restricted sine-Gordon theory, see appendix \ref{rsgappendix}. This theory has no free parameters and no bound states. The case of interest is $p = 4$, for which the restricted theory is supersymmetric.

 
 
The supersymmetric sine-Gordon solitons S-matrix is built in a nice factorized way (arrow  $\raisebox{.5pt}{\textcircled{\raisebox{-.9pt} {4}}}$) from the two S-matrices we just encountered as
\begin{equation*}
S_{\text{SSG}_\text{kinks}}(\theta,\gamma) = S_{\text{SG}_\text{kinks}}(\theta,\gamma) \otimes S_{\text{RSG}_\text{kinks}}^{(p=4)}(\theta)\,,
\end{equation*}
where the SG soliton scattering matrix part takes care of the topological quantum numbers while the RSG matrix deals with the SUSY charges. Just like in SG we can fuse (arrow $\raisebox{.5pt}{\textcircled{\raisebox{-.9pt} {5}}}$) the (supersymmetric) kinks to obtain the S-matrix of the (supersymmetric) breathers, which retains the factorized structure, 
\begin{equation*}
S_{\text{SSG}_\text{breathers}}(\theta,\gamma) = S_{\text{SG}_\text{breathers}}(\theta,\gamma) \otimes S_{\text{RSG}_\text{breathers}}^{(p=4)}(\theta,\gamma)\,.
\end{equation*}
This is the S-matrix at the vertex of figure \ref{fig:masterpiece}. 

Since the fusing momenta depend on $\gamma$, the fusion process introduces a $\gamma$ dependence in the SUSY-related factor. However, this term does \textit{not} correspond to a scattering process in the RSG theory. After all, as said before, the restricted model has no free parameter and no breathers. Nevertheless it is precisely this S-matrix factor by itself that shows up as as the boundary of the purple region in figure \ref{fig:masterpiece}.

The SUSY factor in the SSG 1st breather supermultiplet S-matrix can be deformed into an elliptic integrable S-matrix $S_\text{ED}$ controlled by an extra parameter $\kappa$, arrow  $\raisebox{.5pt}{\textcircled{\raisebox{-.9pt} {6}}}$. This deformation breaks supersymmetry but preserves the $\mathbb{Z}_2$ fermion number symmetry intact. We encounter it as the solid line in the more general $\mathbb{Z}_2$ setup of figure \ref{fig: z2plot}. Finally, it is also possible to deform the SSG 1st breather S-matrix preserving supersymmetry but breaking integrability, see arrow  $\raisebox{.5pt}{\textcircled{\raisebox{-.9pt} {7}}}$ and appendix \ref{nissg}. Such S-matrix, $S_\text{NF}$, describes the full boundary of the space of theories in figure \ref{fig:masterpiece}. It is a curious example of solution which we can find analytically and yet does not obey Yang-Baxter. Would be nice if there was a physical theory which realizes (at least an approximate version) of this S-matrix. 
  
 
The lower dimensional sections of various S-matrix spaces in figures   \ref{fig:masterpiece}, \ref{fig: z2plot} and \ref{z4plot} -- with a vast plethora of very rich S-matrices at their boundary as summarized in figure \ref{web} -- are the main results of this letter. Some of the amazing features in these S-matrix spaces -- such as unitarity saturation -- are now somehow demystified \cite{DualPaperProblem} while others -- such as emerge of factorization or exotic periodicities in the kinematical variables  -- remain as elusive as ever. Would be very interesting to explore other setups with different symmetries and space-time dimensions to better shed light over these puzzles and to best understand how universal they really are. One very concrete avenue for analytic progress is to zoom in on the vertices close to free theories and see if there is still some interesting Lagrangian games to be played a la \cite{NoPP,Bercini}. Would be nice to see if such simple perturbative games, combined with some important bootstrap intuition, could lead to the discovery of new interesting theories. 

\begin{acknowledgments}
We thank Yifei He, Frank Coronado, Kevin Costello, Lorenzo Di Pietro, Davide Gaiotto, Andrea Guerrieri, Martin Kruczenski, Yanyan Li, Guiseppe Mussardo, Joao Penedones, Sasha Zamolodchikov, and specially Lucia Cordova for very useful discussions. Research at the Perimeter Institute is supported in part by the Government of Canada through NSERC
and by the Province of Ontario through MRI. 
This work was additionally supported by a grant from the Simons Foundation (PV: \#488661) and FAPESP grants 2016/01343-7 and 2017/03303-1. 
\end{acknowledgments}

\appendix
\section{Parity and signs}
\label{app:Parity}

In this appendix we formally review the relation between parity and the signs of residues appearing in the S-matrix \cite{Karowski}.

Consider a diagonal scattering process in which a particle $a$ collides against a particle $b$ where both have mass $m$. If $a b$ can form a bound state $c$ of mass $m_c$, the S-matrix $S_{ab}^{ab}$ will contain a pole at $s= m_c^2$:
\begin{equation*}
S_{ab}^{ab}(s\to m_c^2)  \sim -\text{J}_c \frac{{}_{\text{out}}{\<ab|c\>}_{\text{in}}^* {}_{\text{out}}{\<c|ab\>}_{\text{in}}}{s-m_c^2},
\end{equation*}
where $\text{J}_c = m^4/{(2m_c\sqrt{4m^2 - m_c^2})}$ is a Jacobian factor relating the free and interacting parts of the S-matrix. 

We can use a $PT$ transformation to rewrite the first three point function as\footnote{Our discussion is formal because this bound state production process happens for unphysical values of $s$, and so the two-particle states $|ab\>_{\text{in}/\text{out}}$ are schematic.}
\begin{equation*}
{}_{\text{out}}{\<ab|c\>}_{\text{in}} = {}_{\text{in}}{\<ab|c\>}_{\text{out}} \eta_a^* \eta_b^* \sigma_{a b} \eta_c,
\end{equation*}
where $\sigma_{ab} = -1$ if $a$ and $b$ are fermions, one otherwise, and $\eta_x$ is the intrinsic parity of $x$. Therefore the sign of the residue of the $s$-channel pole is given by $-  \eta_a^* \eta_b^* \sigma_{a b} \eta_c$.

Let's compare the general result above with some familiar examples. Recall that bosons may have intrinsic parity $\pm 1$ while Majorana fermions may have intrinsic parity $\pm i$. If we scatter two identical bosons or Majorana fermions, the s-channel residue is always negative, since $ \eta_a^* \eta_b^* \sigma_{a b} = 1 $ in these cases (as is $\eta_c$ from parity conservation).

Next suppose that we scatter a parity even boson and a Majorana fermion of parity $i$. If they form as a bound state a Majorana fermion with the same parity as the external fermion, then the residue in the $s$-channel will be negative as well. On the other hand, if the bound state fermion has parity
$-i$ the residue will be positive. The same would occur when scattering two non-identical even bosons which produce a pseudo-scalar as a bound state. These unusual signs are relevant for the SUSY setup with an anti-fundamental bound state considered in section \ref{susysetup}, see also appendix \ref{SUSYAlgebra}.


\section{Selection rules and crossing}
\label{selection}
In the main text we considered three two-particle scattering scenarios. The SUSY setup, in which we scatter a $\mathcal{N}=1$ supermultiplet against itself, is discussed in detail in appendix \ref{SUSYAlgebra}.  In this appendix we spell out the selection rules imposed by symmetry and the constraints from crossing in the two remaining cases: the $\mathbb{Z}_2$ setup,  where we consider the scattering of two particle states formed out of a degenerate boson and fermion pair, $\left(\phi,\psi\right)$, and the $\mathbb{Z}_4$ setup, where we scatter all two particle states formed out of a particle of unit charge under $\mathbb{Z}_4$, $\mathbf{1}$, together with its antiparticle, $\mathbf{3}$. In all cases we assume that the scattered particles are the lightest in the (gapped) spectrum.

In the $\mathbb{Z}_2$ setup, fermion number symmetry together with parity and time-reversal symmetry impose that the two-to-two S-matrix, in the $\{|\phi \phi\>,|\phi \psi\>,|\psi \phi\>,|\psi \psi\>\}$ basis, is of the form
\begin{equation*}
\mathds{S}_{\mathbb{Z}_2}(\theta) =
\left(\begin{matrix} S_{\phi\phi}^{\phi\phi}(\theta) & 0 & 0 &\displaystyle {S_{\phi\phi}^{\psi\psi}(\theta)}\\
	  0 & {S_{\phi\psi}^{\phi\psi}(\theta)} &  S_{\phi\psi}^{\psi\phi}(\theta) & 0 \\
	 0 & {S_{\phi\psi}^{\psi\phi}(\theta)} &  {S_{\phi\psi}^{\phi\psi}(\theta)} & 0 \\
	 \displaystyle { S_{\phi\phi}^{\psi\psi}(\theta)} & 0 & 0 &\displaystyle {S_{\psi\psi}^{\psi\psi}(\theta)}\end{matrix}\right),
\end{equation*}
where as usual, the rapidity $\theta$ is related to the center of mass energy squared $s$ through $s = 4m^2 \cosh^2(\theta/2)$.

Crossing symmetry relates the scattering amplitudes at different channels through analytic continuation. The diagonal elements are self-crossing and the annihilation and reflection amplitudes, cross into each other: $S_{\phi \phi}^{\psi \psi} (\theta) = S_{\phi \psi }^{\psi \phi } (i\pi-\theta)$.

The bound state spectrum in each setup is implemented through the presence of single poles in each S-matrix element. For example, in the $\mathbb{Z}_2$ setup, assuming the presence of a degenerate boson and fermion pair $(b,f)$ of mass $m_{\text{bs}}$ as bound states, we have
\begin{equation*}
S_{\phi\phi}^{\psi\psi}(s) = - \text{J}_{\text{bs}} \frac{g_{\phi\phi b}g_{\psi\psi b}}{s- m_{\text{bs}}^2} - \text{J}_{\text{bs}} \frac{g_{\phi\psi f}^2}{t- m_{\text{bs}}^2} + \textit{ regular }
\end{equation*}
where \textit{regular} correspond to analytic terms away from the unitarity cuts at $s<0$ or $s>4m^2$ and $\text{J}_{\text{bs}} =m^4/{(2m_{\text{bs}}\sqrt{4m^2 - m_{\text{bs}}^2})}$ is a Jacobian factor.

In the $\mathbb{Z}_4$ setup, the selection rules from charge conservation combined with parity, time-reversal and charge conjugation symmetry constrain the S-matrix to be
\begin{equation}
\mathds{S}_{\mathbb{Z}_4}(\theta) =\left(\begin{matrix} S_{11}^{11}(\theta) & 0 & 0 & \displaystyle {S_{11}^{33}(\theta)}\\
	  0 & {S_{13}^{13}(\theta)}\vspace{0.1cm} &  S_{13}^{31}(\theta) & 0 \\
	 0 & S_{13}^{31}(\theta) &  {S_{13}^{13}(\theta)} & 0 \\
	  \displaystyle { S_{11}^{33}(\theta)} & 0 & 0 & S_{11}^{11}(\theta) \end{matrix} \right)
\label{eq:GeneralZ2S-Matrix}
\end{equation}
in the $\{|11\>,|13\>,|31\>,|33\>\}$ basis. Crossing symmetry acts as $S_{ab}^{cd}(\theta)=S_{a(4-d)}^{c(4-b)}(i \pi - \theta)$ and thus relates the transmission amplitudes as $S_{11}^{11}(\theta)=S_{13}^{13}(i \pi - \theta)$ while the annihilation and reflection amplitudes, ${S_{11}^{33}(\theta)}$ and ${S_{13}^{31}(\theta)}$, are now self crossing symmetric. 

Given a solution of the $\mathbb{Z}_4$ bootstrap setup, i.e. an S-matrix with the correct analytic structure, satisfying unitarity and crossing, one can generate extra solutions by applying independently the following set of transformations:
\begin{align}
\mathds{S}_{\mathbb{Z}_4} &\to  -\mathds{S}_{\mathbb{Z}_4},\nonumber\\
S_{11}^{33} &\to  -S_{11}^{33},\nonumber\\
S_{13}^{31} &\to  -S_{13}^{31},\label{eq:perm} \\
\left(\begin{array}{c} S_{11}^{11} \\ S_{11}^{33} \\ S_{13}^{13} \\ S_{13}^{31} \end{array}\right) &\to \left(\begin{array}{cccc} 0 & 0 & 1 & 0 \\ 0 & 0 & 0 & 1 \\ 1 & 0 & 0 & 0 \\ 0 & 1 & 0 & 0\end{array}\right) \left(\begin{array}{c} S_{11}^{11} \\ S_{11}^{33} \\ S_{13}^{13} \\ S_{13}^{31} \end{array}\right)\nonumber.
\end{align}

The last transformation can be understood as a conjugation of the S-matrix by the operator $\mathds{1} \otimes \mathcal{C}$, each factor acting on a single asymptotic particle, with particles being ordered in two particle states by their rapidities, and $\mathcal{C}$ denoting charge conjugation. The other transformations are trivial or unphysical modifications.


\section{Supersymmetry algebra and representations}
\label{SUSYAlgebra}
The $\mathcal{N}=1$ superalgebra can be written in light-cone coordinates as
\begin{equation*}
\{Q_{-},Q_{+}\}=0\,,\ Q_{+}^2=P_{+}\,, \ Q_{-}^2=P_{-}\,,
\end{equation*}
where $P_{\pm} = me^{\pm\theta}$ are the light-cone momenta.
We realize the algebra as follows
\begin{equation*}
\begin{array}{l}
Q_+|\phi\rangle = \epsilon\sqrt{m} e^{\theta/2}|\psi\rangle \\
Q_+|\psi\rangle = \epsilon^\ast\sqrt{m} e^{\theta/2}|\phi\rangle \\
Q_-|\phi\rangle = \epsilon^\ast\sqrt{m} e^{-\theta/2}|\psi\rangle \\
Q_-|\psi\rangle = \epsilon\sqrt{m} e^{-\theta/2}|\phi\rangle
\end{array}
\end{equation*}
where $\theta$ is the rapidity of the state it acts on and  $\epsilon$ is a phase conventionally chosen to be $\epsilon = e^{-i\pi/4}$ so that crossing is implemented without extra phases, see \cite{nlsm}.

Requiring that the $\mathbb{Z}_2$ S-matrix (\ref{eq:GeneralZ2S-Matrix}) further commutes with the supercharges \cite{Schoutens:1990vb,ahn}, constrains the S-matrix $\mathds{S}_{\text{SUSY}}$ to take the form
\begin{equation}
\begin{array}{c}
\mathds{S}_{\text{SUSY}} =
\left(
\begin{array}{cccc}
S_{\phi\phi}^{\phi\phi} & 0 & 0 & \frac{i(S_{\phi\psi}^{\phi\psi}-S_{\phi\phi}^{\phi\phi})}{\textrm{csch}(\theta/2)} \\
0 & S_{\phi\psi}^{\phi\psi} & \frac{S_{\phi\phi}^{\phi\phi}-S_{\phi\psi}^{\phi\psi}}{\textrm{sech}(\theta/2)} & 0 \\
0 & \frac{S_{\phi\phi}^{\phi\phi}-S_{\phi\psi}^{\phi\psi}}{\textrm{sech}(\theta/2)} & S_{\phi\psi}^{\phi\psi} & 0 \\ 
\frac{i(S_{\phi\psi}^{\phi\psi}-S_{\phi\phi}^{\phi\phi})}{\textrm{csch}(\theta/2)} & 0 & 0 & S_{\phi\phi}^{\phi\phi}-2S_{\phi\psi}^{\phi\psi}
\end{array}
\right)
\end{array}
\label{eq: SUSYgeneral}
\end{equation}
which we can also write as $\mathds{S}_{\text{SUSY}} = \sigma_+ \mathds{T}_+ +\sigma_- \mathds{T}_- $ with 
\begin{equation*}
\mathds{T}_- = 
	\begin{pmatrix}
	\frac{\sinh(\theta/4)^2}{\cosh(\theta/2)} & 0 & 0 & \frac{i\tanh(\theta/2)}{2} \\
	0 & \frac{1}{2} & -\frac{1}{2} & 0 \\
	0 & -\frac{1}{2} & \frac{1}{2} & 0 \\
	\frac{i\tanh(\theta/2)}{2} & 0 & 0 & -\frac{\cosh^2(\theta/4)}{\cosh(\theta/2)}
	\end{pmatrix}
\end{equation*}
and
\begin{equation*}
\mathds{T}_+ =
	\begin{pmatrix}
	\frac{\cosh^2(\theta/4)}{\cosh(\theta/2)} & 0 & 0 & -\frac{i\tanh(\theta/2)}{2} \\
	0 & \frac{1}{2} & \frac{1}{2} & 0 \\
	0 & \frac{1}{2} & \frac{1}{2} & 0 \\
	-\frac{i\tanh(\theta/2)}{2} & 0 & 0 & -\frac{\sinh(\theta/4)^2}{\cosh(\theta/2)}
	\end{pmatrix}\,.
\end{equation*}

The tensors $\mathds{T}_i$ are invariant under supersymmetry and are constructed such that
\begin{equation*}
\mathds{T}_i(\theta)\mathds{T}_j(-\theta) = \delta_{ij}\mathds{P}_i(\theta)\,,
\end{equation*}
where $\mathds{P}_i$ are orthonormal projectors:
\begin{equation*}
\begin{array}{l}
\mathds{P}_i(\theta)\mathds{P}_j(\theta)=\delta_{ij}\mathds{P}_i(\theta)\,,\\
\mathds{P}_-(\theta)+\mathds{P}_+(\theta)=\mathds{1}\,.
\end{array}
\end{equation*}

Using these properties we can simply write 
\begin{equation*}
S(\theta)S(-\theta) = |\sigma_-(\theta)|^2\mathds{P}_-(\theta)+|\sigma_+(\theta)|^2\mathds{P}_+(\theta)\,.
\end{equation*}
In sum, the advantage of  this parametrization is that it trivializes unitarity to 
\begin{equation*}
|\sigma_+(\theta)|^2\leq 1\;\;\textrm{and}\;\;|\sigma_-(\theta)|^2\leq 1\,.
\label{eq:SigmaUnitarity}
\end{equation*}
At this point unitarity is cast in the same spirit of previous S-matrix bootstrap works \cite{lucia,Guerrieri:2018uew}. As in those works, we are splitting the symmetry group into irreducible representations associated with the projectors $\mathds{P}_+$ and $\mathds{P}_-$ corresponding to the fundamental and anti-fundamental representations of the supersymmetry algebra. Within each channel, unitarity is as straightforward as for a single component scattering. 

To put a bound state excitation in a particular representation we must put a single pole in the correspondent $\sigma_i$ function, or conversely, require that the residue of the other representation is zero. If we let $\theta_*=i\gamma$ be the position of the bound state pole in the $\theta$-plane, we have the following relations between the coupling strenghts\footnote{We can also arrive at the coupling relations by writing the residues as three point functions and use supersymmetric Ward identities.}:
\begin{equation}
\begin{array}{l}
\text{fundamental}: g_{\phi\phi b}^2 = g_{\phi\psi f}^2\left(1 + \sec\left(\tfrac{\gamma}{2}\right)\right), \\ \\
\text{anti-fundamental}: g_{\phi\phi b}^2 = g_{\phi\psi f}^2\left(-1 + \sec\left(\tfrac{\gamma}{2}\right)\right).
\end{array}
\label{eq:SUSYResidues}
\end{equation}

The difference in the signs of residues in each case can be interpreted as a difference in the parities of the bound states, see appendix \ref{app:Parity}. In both scenarios the boson is parity even, but the fermionic bound state differs. It has the same parity as the external fermion when the multiplet is in the fundamental representation and the opposite parity when it is in the anti-fundamental representation.

\section{Exact S-matrices} 
\label{exact}

In this appendix we briefly review the exact S-matrices and related field theories showing up in this work. These include the regular sine-Gordon model (SG) \cite{sgkinks}, the supersymmetric sine-Gordon model (SSG) \cite{ahn}, the restricted sine-Gordon model (RSG) \cite{bernard, rsgSmirnov}, Zamolodchikov's $\mathbb{Z}_4$ S-matrix \cite{z4paper} and, so far as we are aware, a novel elliptic deformation of the SSG breathers S-matrix~\cite{Paper4}, and a new non-factorizable deformation of SSG. The relations between the various S-matrices are summarised in section \ref{relations}.

\subsection{Sine-Gordon}\label{sgappendix}

\begin{figure}[t!]
	\centering 
	\includegraphics[width=\linewidth]{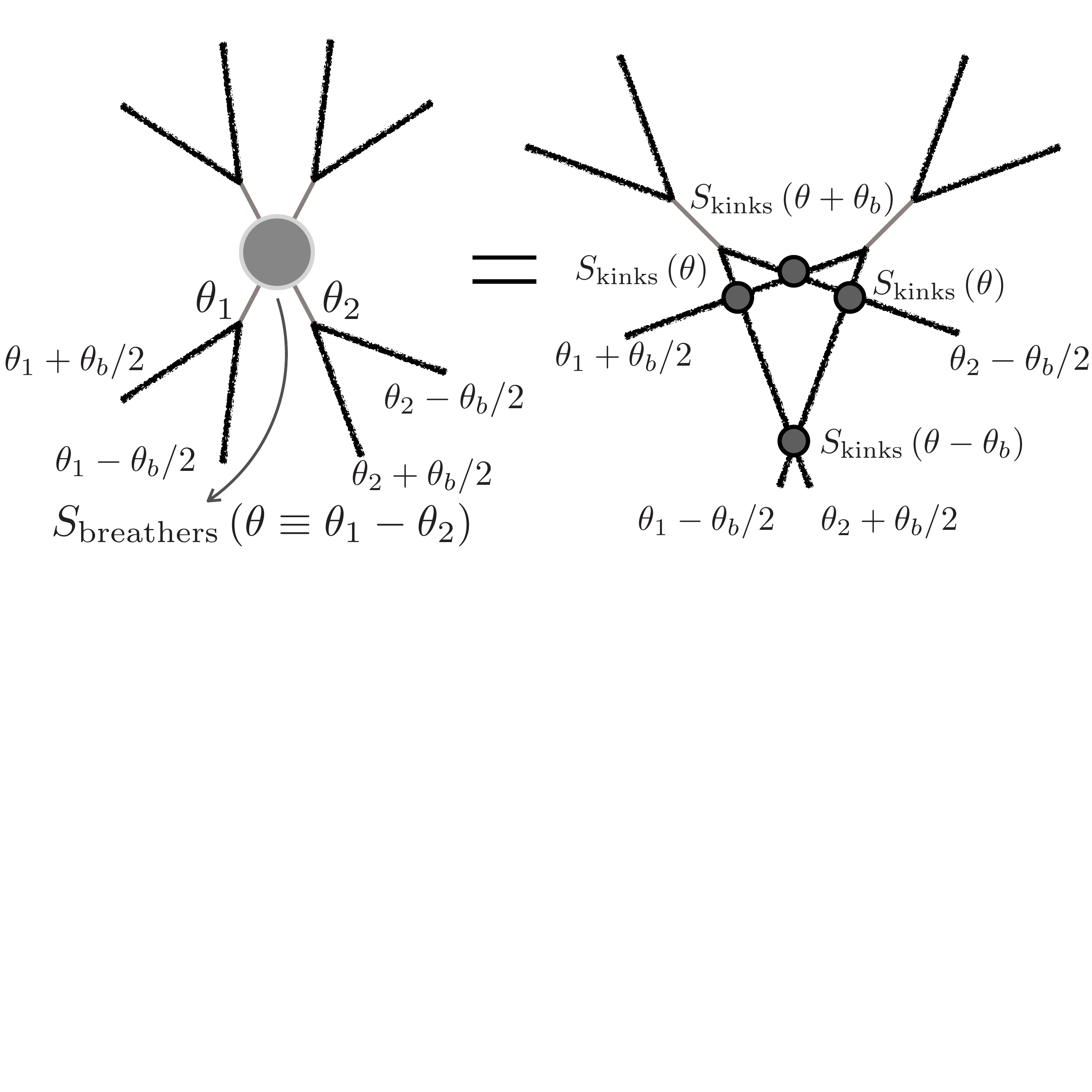}
	\vspace{-4cm}
\caption{Suppose we scatter two pairs of kinks, the constituents of each pair having the precise relative rapidity  $\theta_b$ so as to form a breather. We then let this bound states collide and later decay. That will correspond to a double pole of the $4\to4$ kink S-matrix whose coefficient is proportional to the breather-breather S-matrix of the theory. This is the process described in the left hand side of the picture. From integrability, we can rearrange the incoming wave packets so that the kinks scatter before fusing into bound states, as in the right hand side of the figure. In this way, we relate the breathers S-matrix to a factorised product of four two-to-two kinks S-matrix. For simplicity, we omitted quantum numbers  that would be relevant in the sine-Gordon or supersymmetric sine-Gordon theories, such as topological or SUSY charges. The fusing angle is fixed both in SG and SSG to be $\theta_b = i \gamma$.}\label{fusion}
\end{figure}

We begin with the regular sine-Gordon theory, whose action is
\begin{equation*}
A_{\text{SG}} = \frac{(\gamma + \pi)}{8\pi\gamma}\int d^2x\left(  \frac{\partial_\mu \phi \,\partial^\mu \phi}{2} + m^2 \left(\cos{\phi}-1\right)\right),
\la{sGlag}
\end{equation*}
where $\gamma$ is the effective coupling. For $\gamma \geq \pi$ the spectrum consists of solitons $\{|+\rangle, |-\rangle\}$ carrying $U(1)$ topological charges. Their exact scattering \cite{sgkinks} matrix $S_{\textrm{SG}_\textrm{kinks}} (\theta)$, in the \mbox{$\{|++\>, |+-\>, |-+\>, |--\>\}$} basis, is equal to 
\begin{equation}
\begin{array}{l}
\mathcal{U}(\theta)\!\times \!
	\begin{footnotesize}
	\left(\!
		\begin{array}{cccc}
		\sinh\frac{\pi \left(i \pi - \theta\right)}{\gamma} & 0 & 0 & 0 \\
		0 & \sinh\frac{\pi\theta}{\gamma}  &  i \sin\frac{\pi^2}{\gamma}& 0 \\
		0 & i \sin\frac{\pi^2}{\gamma} &  \sinh\frac{\pi\theta}{\gamma} & 0 \\
		0 & 0 & 0 &  \sinh\frac{\pi \left(i \pi - \theta\right)}{\gamma}
		\end{array}\!
	\right)	
	\end{footnotesize}
\end{array} \label{sgkinks}
\end{equation}
where $\mathcal{U}(\theta) = \frac{\Gamma\big(\tfrac{\pi}{\gamma}\big)\Gamma\big(1 + i \tfrac{\theta}{\gamma}\big) 
\Gamma\big(1 - \tfrac{\pi}{\gamma} - i \tfrac{\theta}{\gamma}\big)}{i \pi}\prod\limits_{n=1}^\infty \frac{F_n \left(\theta\right)F_n \left(i \pi - \theta\right)}{F_n \left(0\right)F_n \left(i\pi \right)}$ with $F_n (\theta) = \frac{\Gamma\left(\frac{2n\pi + i\theta}{\gamma}\right) \Gamma\left(1 + \frac{2n\pi+i\theta}{\gamma}\right)}{\Gamma\left(\frac{(2n +1)\pi+ i\theta}{\gamma}\right) \Gamma\left(1 + \frac{(2n -1)\pi+i\theta}{\gamma}\right)}$.
The S-matrix~(\ref{sgkinks})  corresponds to the green edge along the boundary of the $\mathbb{Z}_4$ symmetric S-matrices of figure \ref{z4plot}. The edge is parameterized by $\gamma \in (\pi,\infty)$, with $\gamma = \pi$ corresponding to free field theory.

For $\gamma < \pi$ the solitons can form bound-states called breathers. In integrable theories, the bound states S-matrix can be obtained from the fusion of the S-matrices of their constituents, figure \ref{fusion}. For the lightest  breather of sine-Gordon this gives
\begin{equation*}
S_{\textrm{SG}_\textrm{breathers}}\left(\theta\right) =\frac{\sinh{\theta} + i \sin{\gamma}}{\sinh{\theta} - i \sin{\gamma}}\,, 
\label{sgbreathers}
\end{equation*}
which appeared in the S-matrix bootstrap context in \cite{Paper2}. There it was shown (both analytically and numerically) that this S-matrix has the biggest coupling between the external particles (lightest breather) and their bound state (second-lightest breather).

\subsection{Restricted sine-Gordon}
\label{rsgappendix}

 \begin{figure}[t!]
	\centering 
	\includegraphics[width=\linewidth]{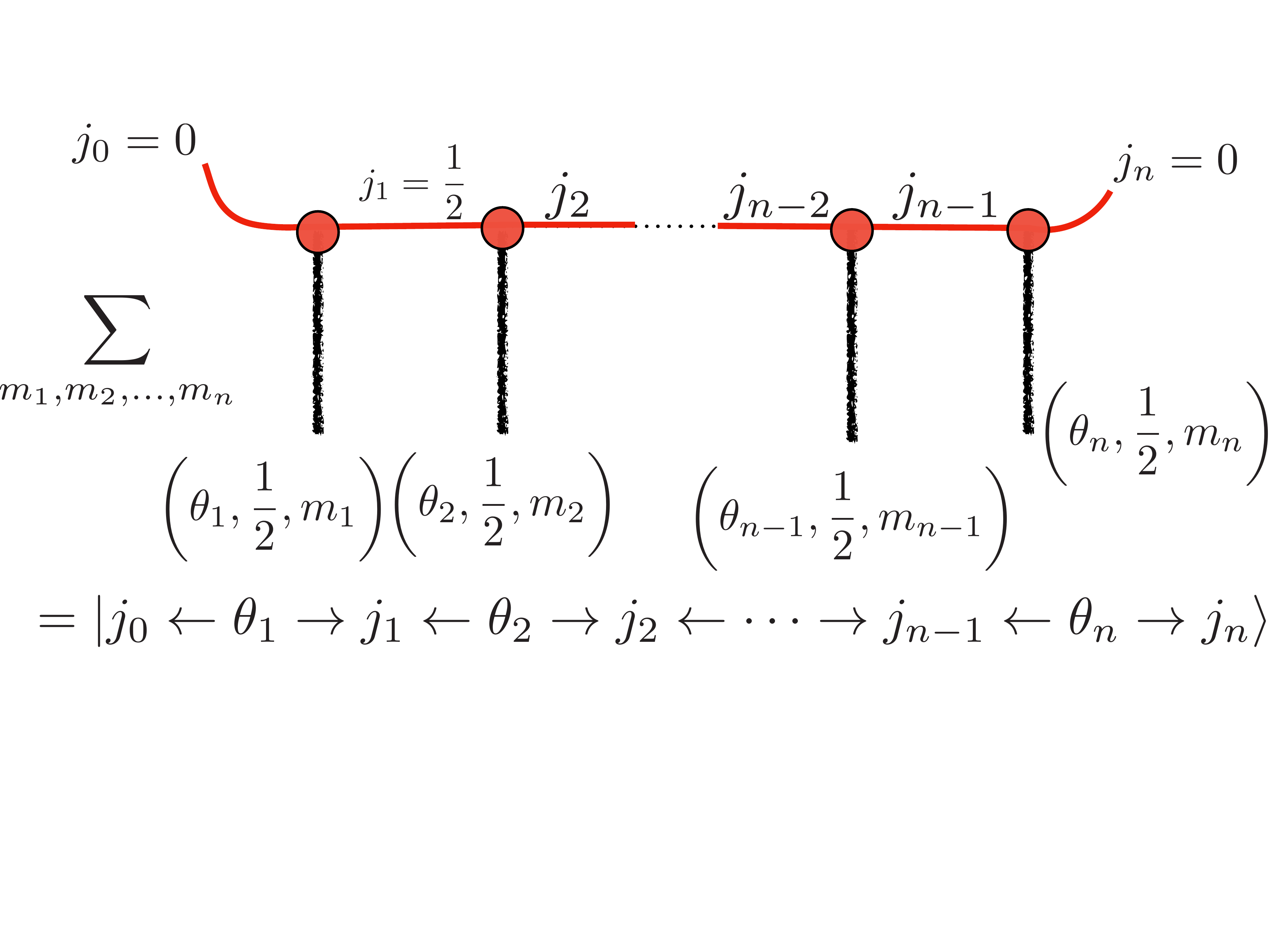}
	\vspace{-2cm}
\caption{Consider an $n$-soliton state in sine-Gordon theory.  Each soliton carries a rapidity $\theta_i$ and a spin 1/2 index $m_i$. To construct a basis of $U_q(\mathfrak{sl}(2))$ invariant states we order the solitons in a comb-like structure and project into intermediate $U_q(\mathfrak{sl}(2))$ partial-waves at each vertex by contracting the incoming $U_q(\mathfrak{sl}(2))$ indices with some appropriately normalised $3j$-symbols, see \cite{rsgSmirnov} for details. Here we omit and sum over the spin indices of the intermediate states, as well as sum over external indices $m_i$. The state is invariant if we require that the final symbol projects the state into the spin 0 representation. The invariant subspace is spanned by different decomposition histories $(0, 1/2, j_2,\dots,j_{n-2},1/2,0)$. In this basis, it is useful to think of each rapidity $\theta_i$ as carrying two quantum numbers $\left(j_{i-1},j_i\right)$, see figure \ref{rsgmatrix}.} 
\label{rsgstate}
\end{figure} 

The sine-Gordon theory possess  $U_q(\mathfrak{sl}(2))$ quantum symmetry with $q= -e^{-i \pi^2/\gamma}$. The physics of the model is drastically modified when $q$ is a root of identity, {\it i.e.}, for $\gamma=\pi p$, with $p\geq 3$ and $p\in\mathbb{Z}$. For this values some multi-soliton states decouple and the spectrum can be restricted. It is then useful to introduce a new basis of particles, as described in figure \ref{rsgstate}, each carrying a rapidity and two $U_q(\mathfrak{sl}(2))$ spin quantum numbers. 

The S-matrix between these new excitations is obtained from the fundamental solitons S-matrix (\ref{sgkinks}) through an interaction-round-a-face to vertex transformation. For a given $p$, the RSG kinks S-matrix is defined by
\begin{align}
&|j_0 \leftarrow \theta_1 \rightarrow j_1 \dots j_{i-1}\leftarrow \theta_i \rightarrow j_i \leftarrow \theta_{i+1} \rightarrow j_{i+1}  \dots  j_n \rangle = \nonumber\\&
\sum_{j_i}S_{\textrm{RSG}_\textrm{kinks}}^{(p)} \left(\left.{\begin{array}{cc}
	j_{i-1} & j'_i \\
	j_{i} & j_{i+1} \\
	\end{array}}
\right|\theta\equiv\theta_i-\theta_{i+1}\right) \label{rsGsmatrix} \\
&|j_0 \leftarrow \theta_1 \rightarrow j_1 \dots  j_{i-1}\leftarrow \theta_i \rightarrow j_i \leftarrow \theta_{i+1} \rightarrow j_{i+1} \dots j_n \rangle. \nonumber
\end{align}

\begin{figure}[t!]
	\centering 
	\includegraphics[width=\linewidth]{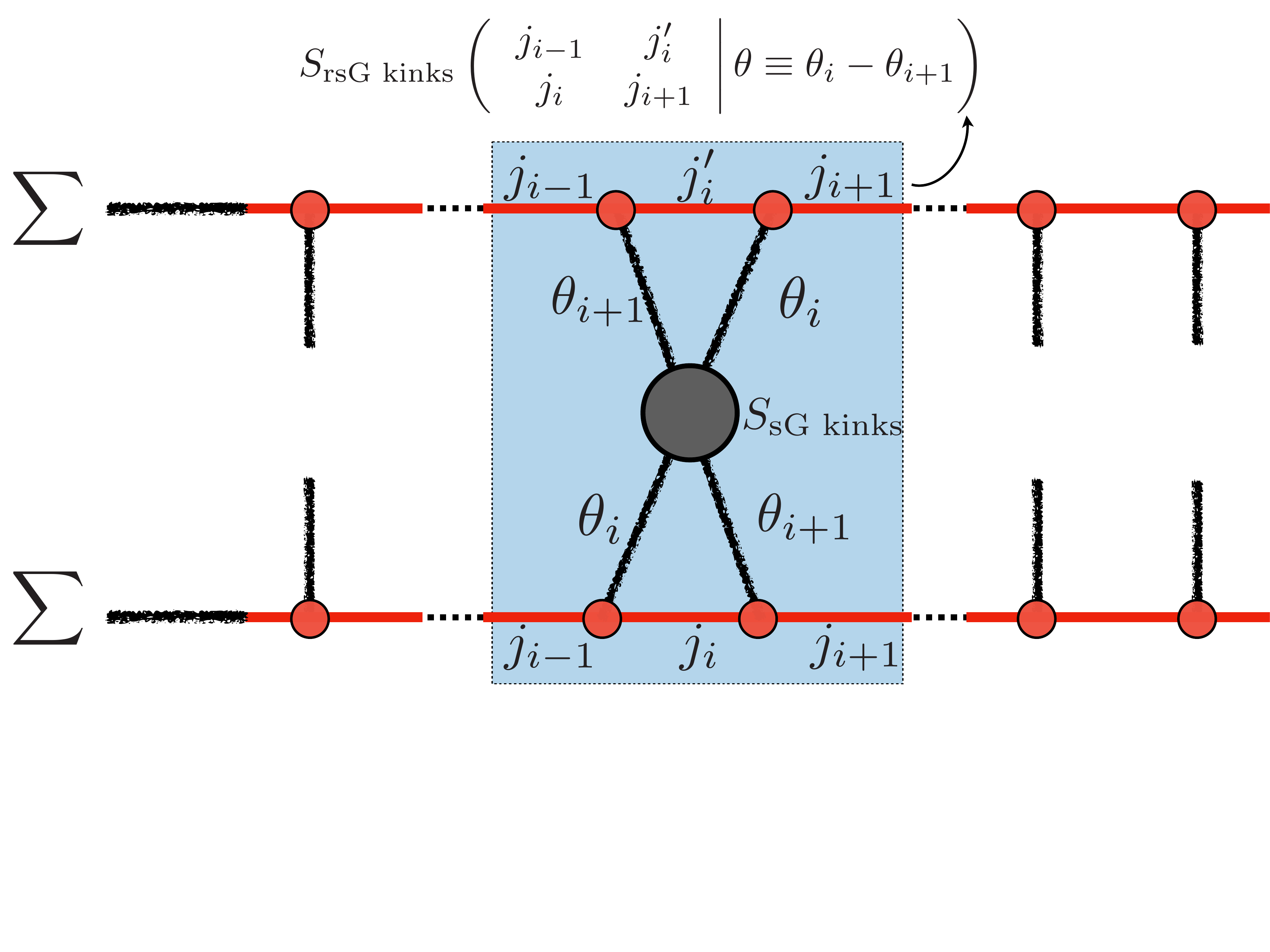}
	\vspace{-2cm}
\caption{Each RSG kink carries a rapidity $\theta_i$ and quantum numbers $\left(j_{i-1},j_i\right)$ associated to the neighbouring intermediate spins. Their S-matrix is defined by equation (\ref{rsGsmatrix}) in terms of the states described in figure \ref{rsgstate}, and its relation to the regular SG kinks' S-matrix is illustrated above. As before, spin indexes are summed over and ommited. Note that the incoming and outgoing kinks must share the $j_{i-1}$ and $j_{i+1}$ quantum numbers: this is an IRF type S-matrix.}
\label{rsgmatrix}
\end{figure} 

As explained in figure \ref{rsgmatrix}, one can then determine this S-matrix in terms of $S_{\text{SG}_\text{kinks}}$ and the $U_q(\mathfrak{sl}(2))$ $3j$-symbols \cite{bernard}, to be
\begin{equation*}
\begin{array}{l}
S_{\text{RSG}_\text{kinks}}^{(p)} 
\left(\left.
\begin{array}{cc}
a&b\\
c&d\\
\end{array}\right|
\theta\right) = \frac{\mathcal{U}(\theta)}{2}\left(\frac{[2a+1][2b+1]}{[2c+1][2d+1]}\right)^{\frac{i\theta}{2\pi}}\mathcal{R}_{cd}^{ab}\\ \\
\mathcal{R}_{cd}^{ab} = \sqrt{\frac{[2a+1][2b+1]}{[2c+1][2d+1]}}\sinh\left(\frac{\pi\theta}{\gamma}\right)\delta_{ad}+\sinh{\left(\frac{i\pi-\theta}{\gamma}\right)}\delta_{cb}\,, \\ \\
\[x\] = \frac{q^x-q^{-x}}{q-q^{-1}}\,,
\end{array}
\end{equation*}
where $\mathcal{U}(\theta)$ has the same form as in $S_{\text{SG}_\text{kinks}}$.

As a consistency check on the restriction one can use this explicit form to verify that the RSG scattering amplitudes vanish whenever $a,c,d \leq p/2 - 1< b $.  For $p=4$ the quantum group charges act on the scattering states described in figure \ref{rsgstate} as $\mathcal{N}=1$ supersymmetry\footnote{Strictly speaking one has to do a change of basis on this states to define a canonical basis that transforms appropriately under supersymmetry as detailed in \cite{ahn}. }.

\subsection{Zamolodchikov's $\mathbb{Z}_4$ S-matrix}\la{sec:exactz4matrix}

It turns out that for $\gamma>\pi$ the sine-Gordon kinks' S-matrix admits a one-parameter deformation which preserves integrability. This is the $\mathbb{Z}_4$-symmetric elliptic S-matrix of Zamolodchikov \cite{z4paper}, and is the basic building block to construct the full boundary of the space considered in figure \ref{z4plot}. It describes the two-to-two scattering in a theory with two particles $\{|1\rangle,|3\rangle\}$. These form a particle-antiparticle pair with charges one and three under $\mathbb{Z}_4$, respectively. The explicit S-matrix $S_{\mathbb{Z}_4}(\theta)$  in the $\{|11\>, |13\>, |31\>, |33\>\}$ basis is equal to 
\begin{equation*}
\begin{array}{l}
	 \mathcal{U}(\theta)\times 
	\left(
		\begin{array}{cccc}
		\frac{\text{sn}_{(\pi+i \theta)\alpha}}{\text{sn}_{\pi\alpha}} & 0 & 0 & \frac{\text{sn}_{(\pi+i \theta)\alpha}\text{sn}_{i\theta\alpha}}{k}\\
		0 & -\frac{\text{sn}_{i\theta\alpha}}{\text{sn}_{\pi\alpha}} & 1 & 0 \\
		0 & 1 & -\frac{\text{sn}_{i\theta\alpha}}{\text{sn}_{\pi\alpha}} & 0 \\
		\frac{\text{sn}_{(\pi+i\theta)\alpha}\text{sn}_{i\theta\alpha}}{k} & 0 & 0 & \frac{\text{sn}_{(\pi+ i\theta)\alpha}}{\text{sn}_{\pi\alpha}}
		\end{array}
	\right)
\end{array}
\label{z4matrix}
\end{equation*}
where $k=-1/\sqrt{\kappa}$ and where here 
\begin{equation*}
\begin{normalsize}
\begin{array}{l}
\mathcal{U}(\theta) = 
\exp\sum\limits_{n=1}^\infty \frac{4\sinh^2\left(2\pi n \left(\pi - \beta\right)/\beta'\right)\sin\left( 2 \pi n (i \pi - \theta)/\beta' \right)}{n \csc\left(2\pi n\theta/\beta'\right) \sinh{\left(4 \pi n \beta/\beta' \right)} \cosh{\left(2 \pi^2 n/\beta'\right)}}\,, \\ \\
\beta = 2 K(\kappa)/\alpha\,, \;\; \beta' = 2K(1-\kappa)/\alpha\,, \;\;\text{sn}_{x} \equiv \text{sn}(x,\kappa)\,,
\end{array}
\end{normalsize}
\end{equation*}
with $K(\kappa)$ denoting the complete elliptic integral of the first kind and sn the Jacobi elliptic sine and $\alpha = \pi/\gamma$. 

From real analyticity the deformation parameter $\kappa$ must takes values in $\[0,1\)$ while $\alpha$ must be either purely imaginary or real. 
In the $\kappa \to 0$ limit we recover the sine-Gordon kinks S-matrix.
It turns out that due to periodicity on the $\theta$-plane the coupling value must be further constrained by $\alpha \in i \left(0, \frac{K(1-\kappa)}{\pi}\right)$ or $\alpha \in   \left(0, \frac{2 K(\kappa)}{\pi}  \right)$ to prevent unphysical poles from coming into the physical sheet. Applying transformations (\ref{eq:perm}) to Zamolodchikov's $\mathbb{Z}_4$ S-matrix the full boundary of the space described in figure \ref{z4plot} is obtained.

\subsection{Minimal supersymmetric sine-Gordon} \la{sec:ssg}

The $\mathcal{N}=1$ supersymmetric sine-Gordon action is given by
\begin{align}
A_{\text{SSG}} & = \frac{(\gamma + 2\pi)}{4\pi\gamma}\int d^2x \biggl( \frac{\partial_\mu \phi \,\partial^\mu \phi}{2} + \frac{i}{2} \bar{\psi}  \slashed{\partial} \psi + \nonumber\\ 
& + \frac{m^2}{4} \cos{\phi}^2 -\frac{m}{2}\bar{\psi}\psi\cos{\phi} \biggr).\nonumber
\end{align}

Just as sine-Gordon, for $\gamma<\pi$, the spectrum contains bound states (breathers). And by the same process of fusion, described in figure \ref{fusion}, we obtain the S-matrix of the lightest breather supermultiplet, \cite{ahn}. In the $\{|\phi\phi\>, |\phi\psi\>, |\psi\phi\>, |\psi\psi\>\}$ basis it is given by 
{\small
\begin{align} \la{ssGbsmatrix}
& S_{\text{SSG}_{\text{breather}}}\left(\theta\right) = S_{\text{SG}_{\text{breather}}}\left(\theta\right)\mathcal{U}\left(\theta\right)\times \\
&\begin{pmatrix}
\frac{2i \sin\left(\gamma/2\right)}{\sinh\left(\theta\right)}{\small+1} & 0 & 0 &\frac{\sin\left(\gamma/2\right)}{
	\cosh\left(\frac{\theta }{2}\right)} \\
0 & 1 & \frac{i \sin\left(\gamma/2\right)}{ \sinh\left(\frac{\theta }{2}\right) }& 0 \\  
0 & \frac{i \sin\left(\gamma/2\right)}{ \sinh\left(\frac{\theta }{2}\right) } & 1 & 0  \\
\frac{\sin\left(\gamma/2\right)}{
	\cosh\left(\frac{\theta }{2}\right)}  & 0 & 0 & \frac{2i \sin\left(\gamma/2\right)}{\sinh\left(\theta\right)}{\small -1}
\end{pmatrix}\nonumber
\end{align}}
where
\begin{align*}
& \mathcal{U}\left(\theta\right) = \\
& \biggl[ \frac{\Gamma\left(-i \theta/2\pi \right)}{\Gamma\left(1/2 -i \theta/2\pi \right)}\prod_{n=1}^\infty\biggl(\frac{\Gamma\left(\gamma/2\pi  - (i \theta/2\pi) + n \right)}{\Gamma\left(\gamma/2\pi  - (i \theta/2\pi) + n + 1/2\right)}\nonumber \times\\
&\frac{\Gamma\left(-\gamma/2\pi  - (i \theta/2\pi) + n -1\right) \Gamma^2\left(-(i\theta/2\pi) + n -1/2 \right)}
{\Gamma\left(-\gamma/2\pi  - (i \theta/2\pi) + n -1/2\right) \Gamma^2\left(-(i\theta/2\pi) + n -1 \right)}\biggr)\biggr]\nonumber\times\\
&[\theta\to i\pi-\theta]\nonumber
\end{align*}
\normalsize

The poles in the lightest breather S-matrix correspond to the second-lightest breather supermultiplet of the spectrum.


It turns out that the supersymmetric sine-Gordon S-matrix is completely fixed by supersymmetry and Yang-Baxter  \cite{Schoutens:1990vb,nlsm}. Indeed, requiring that the general SUSY S-matrix (\ref{eq: SUSYgeneral}) satisfies the Yang-Baxter condition implies that $S_{\phi\phi}^{\phi\phi}/S_{\phi\psi}^{\phi\psi} = 1 + i \alpha/\sinh\theta/2$ with $\alpha$ a constant. The overall factor is then fixed by unitarity up to CDD ambiguities. Furthermore, by requiring that the residues in different matrix elements are consistent with a bound state in the fundamental representation we fix $\alpha$ and obtain the matrix structure of the SSG breathers S-matrix (\ref{ssGbsmatrix}). For a bound state in the anti-fundamental representation, the S-matrix is similarly fixed to be $S_{\text{SSG}_\text{breathers}}$ analytically continued to $\gamma\to\gamma+2\pi$.

\subsection{Elliptic deformation of the supersymmetric sine-Gordon}\label{sec:ellipticdef}

In \cite{Paper4} a Yang-Baxter preserving but supersymmetry breaking deformation of (\ref{ssGbsmatrix}) was obtained. The S-matrix is
\begin{equation*}
\begin{array}{l}
	S_{\text{ED}}(\theta) = S_{\text{SG}_{\text{breathers}}}(\theta)\mathcal{U}(\theta)\times \\ \\
	\left(
		\begin{array}{cccc}
		\frac{\text{dn}_{\theta\omega} \text{sn}_{i \gamma\omega}}{  \text{cn}_{\theta\omega} \text{sn}_{\theta\omega}}+\text{dn}_{i \gamma\omega} & 0 & 0 & \frac{\text{dn}_{\theta\omega} \text{sn}_{i \gamma\omega}}{  \text{cn}_{\theta\omega}}\\  
 0 & 1 & \frac{\text{sn}_{\gamma\omega}}{  \text{sn}_{\theta\omega}} & 0\\ 
 0 & \frac{\text{sn}_{\gamma\omega}}{  \text{sn}_{\theta\omega}} & 1 & 0 \\  
 \frac{\text{dn}_{\theta\omega} \text{sn}_{i \gamma\omega}}{  \text{cn}_{\theta\omega}} & 0 & 0 & 
\frac{\text{dn}_{\theta\omega} \text{sn}_{i \gamma\omega}}{  \text{cn}_{\theta\omega} \text{sn}_{\theta\omega}}-\text{dn}_{i \gamma\omega}
\end{array}
\right)
\end{array}
\label{ed}
\end{equation*}
where
\begin{equation*}
\begin{array}{l}
\mathcal{U}(\theta) =
-i \sinh\left(\theta\right) \text{exp} \left(\int_{-\infty}^{\infty} \frac{d\theta'}{2\pi i}\frac{\log\left(g(\theta')/\sinh(\theta')^2\right)}{\sinh\left(\theta - \theta' + i \epsilon\right)} \right)\,\\ \\
\omega = -\frac{i}{\pi} K(\kappa)\,,\;\; g(\theta) = 1 - \frac{\text{sn}(i\gamma\omega,\kappa)^2}{\text{sn}(\theta\omega,\kappa)^2}\,, \\ \\
\text{sn}_x = \text{sn}(x,\kappa)\,,\;\; \text{dn}_x = \text{dn}(x,\kappa)\,,\;\;\text{cn}_x = \text{cn}(x,\kappa)\,.
\end{array}
\end{equation*}

The deformation parameter $\kappa$ is constrained to the interval $\left(-\infty, 1\right)$ due to real analyticity, with the SSG breathers S-matrix being recovered in the $\kappa \to 0$ limit. The residues of this S-matrix as a function of $\kappa$ correspond to the solid line in figure \ref{z4plot}.

\subsection{Non factorizable deformation of supersymmetric sine-Gordon}\label{nissg}
Following the steps described in appendix \ref{App:AnalyticalBound} we were able to obtain an analytical expression for the supersymmetric S-matrices that lies along boundary of the space described in figure \ref{fig:masterpiece}:
\begin{equation}
\begin{array}{l}
S_{\textrm{NF}}(\theta) = S_{\text{SSG}_{\text{breathers}}}\left(\theta\right)\mathcal{U}(\theta)\times \\ \\
\left(
\begin{array}{cccc}
r(\theta)& 0 & 0 & \frac{i(1-r(\theta))}{\textrm{csch}(\theta/2)} \\
0 & 1 & \frac{r(\theta)-1}{\textrm{sech}(\theta/2)} & 0 \\
0 & \frac{r(\theta)-1}{\textrm{sech}(\theta/2)} & 1 & 0 \\ 
\frac{i(1-r(\theta))}{\textrm{csch}(\theta/2)} & 0 & 0 & r(\theta)-2
\end{array}
\right)
\end{array}\,,
\label{eq:NIDeformation}
\end{equation}
where
\begin{align*}
&\mathcal{U}(\theta)=
{\small \pm \left(\frac{\sinh\theta-i\sqrt{t}}{\sinh\theta+i\sqrt{t}}\right)^{\Theta(t)}} \\
&{\small \times \exp\left(-\int_{-\infty}^\infty\frac{d\theta^\prime}{2\pi i}\frac{\log\left(1-\cosh^2\left(\theta^\prime/2\right)(1-r(\theta^\prime))^2\right)}{\sinh(\theta^\prime-\theta+i\epsilon)}\right)\,,}\nonumber 
\end{align*}
with $r(\theta)$ being the ratio function found in (\ref{eq:FinalRatio}).

By varying the parameter $t \in \mathbb{R}$ and the overall signs in $\mathcal{U}$ we parametrize the full boundary of figure \ref{fig:masterpiece}. The CDD-zero only makes sense for positive $t$. Negative values of this parameter would break real analyticity and introduce poles in physical of scattering energies, hence the presence of the step function. The reader can see that $t=0$ yields the SSG model.

\section{More on the ratio function}
\label{App:AnalyticalBound}
The S-matrices on the boundary of the supersymmetric bootstrap are unitary. This constrains the S-matrices to be on the form (\ref{eq:NIDeformation}) with the condition that
\begin{equation}
r(\theta) + r(-\theta) =  2
\label{eq:AnalyticalUnitarity2}
\end{equation}
where $r$ is simply the ratio between the two independent S-matrices elements ($r= S_{\phi\phi}^{\phi\phi}/S_{\phi\psi}^{\phi\psi}$).

Using crossing symmetry in the relation above, is easy to see that this ratio function is  $2\pi i$-periodic. This allows us to look only at two sheets of the $\theta$-plane. It turns out that on the boundary the ratio has a very simple analytical structure. In the first sheet the numerical solutions have a pole at $ \theta=i\delta_1$, a zero at $\theta = i\delta_2$ and their corresponding crossing symmetric partners. On the second sheet they have the same poles plus an extra zero at $\theta=i\pi+\delta_3$ and its crossing symmetric partner\footnote{Numerically we observe that the $\delta$'s belong to $\left(0,\pi\right)$ or \\$\{\pi/2 + i \tau | \tau \in \mathbf{R}\}$.}.

We start with an ansatz manifestly crossing symmetric and with the correct analytic structure
\begin{equation*}
\begin{array}{l}
r(\theta) = A \left(\frac{\sinh\left(\frac{1}{2}(\theta-i\delta_2)\right)\sinh\left(\frac{1}{2}(\theta-i\delta_3)\right)}{\sinh\left(\theta-i\delta_1\right)}\right)\times\left(\delta_j \to \pi - \delta_j\right)\ .
\end{array}
\end{equation*}

The unitarity constraint (\ref{eq:AnalyticalUnitarity2}) fixes $\delta_3$ as a function of $\delta_1$ and $\delta_2$, leaving two free parameters. Let $\theta_*=i\gamma$ be the position of the bound state pole. The fact that the residues of the supersymmetric S-matrix elements are related by (\ref{eq:SUSYResidues}) gives another constraint,
\begin{equation*}
r(i\gamma) =
 1 \pm \sec(\gamma/2)\,,
\label{eq:ResidueAnsatz}
\end{equation*}
where the sign reflects which representation one chooses. This relation fixes $\delta_2$ as a function of $\delta_1$.

At the SSG point (or at its equivalent for the anti-fundamental representation) we have $\delta_1 =0$ and therefore
\begin{equation*}
r(\theta)|_{\delta_1=0} =
 1 \pm \frac{2i\sin(\gamma/2)}{\sinh\theta}\,,
\end{equation*}
which fixes the overall constant $A$.

The final solution then depends on two parameters: $t=\sin^2\delta_1$ and $\gamma$ that determines the bound state mass,
\begin{equation}
r(\theta) = 1 \pm i\left(2\sin\left(\frac{\gamma}{2}\right)-\frac{t}{\sin{\gamma}\cos\left(\frac{\gamma}{2}\right)}\right)\frac{\sinh(\theta)}{t+\sinh(\theta)^2}\,.
\label{eq:FinalRatio}
\end{equation}
Note that when $t=0$ we recover the SSG model as expected and when $t=\sin(\gamma)^2$ we reach the free theories points.

\section{Numerics} \label{appendixNumerics}


Our numerics follow verbatin the algorithms in \cite{Paper2,Paper4,Martin}. In short we first propose a very general ansatz for the S-matrix elements in terms of a large linear combination of basis functions as (here the index $a$ labels all possible scattering channels)
\begin{equation*}
S_a(s)=\texttt{poles}_a+ \texttt{regular}_a= \sum_{n=1}^N c_a^{(n)} f_n(s) \,.
\end{equation*}
with $N$ as large as our computers allow. What are these functions $f_n$? They can be any basis which spans the full space of possible S-matrices -- with their required analytic properties -- as $N\to \infty$. Common examples are Fourier series, Taylor expansions, (discretized) dispersion relations etc. We use the latter for the plots in this letter. 

Note that both the S-matrix elements at some off-shell value, $S_a(s_*)$, as well as the residues of the poles of the S-matrix elements are then explicit linear combinations of the $c_a^{(n)}$. These linear combinations are what we want to maximize or minimize to determine the boundary of the allowed S-matrix space. 

So all we have to do is to maximize these linear combinations subject to the two relevant physical constraints which are crossing and unitarity. Crossing $S_a(4-s)=C_{ab} S_b(s)$ is a simple linear constraints on the $c_a^{(n)}$'s. We can use it to simply eliminate some of these constants in terms of the others. Unitarity is more interesting. In terms of probability conservation it reads as $|S_a(s)|^2\le 1$ for any real $s$ above the two-particle production threshold. This condition can be trivially linearized as the statement that the matrix 
\begin{equation*}
\( \begin{array}{cc} 1 & S_a(s) \\ S_a(s)^* & 1 \end{array} \)
\end{equation*}
is positive semi-definite in that same range of $s$. In practice we impose this condition in a grid in $s$ starting from threshold and going to some large energy value. For each $s$ we get a positive semi-definite condition, all of which linear in all the parameters $c_a^{(n)}$. Hence our maximization problem is nothing but what is called a semidefinite programming (SDP) problem for which we can use the very powerful \texttt{sdpb} software developed by Simons-Duffin \cite{duffin}. That is what we did.

\newpage
\bibliographystyle{apalike2}
\bibliography{bibli.bib}

\end{document}